\documentclass[pre,aps,floatfix,twocolumn, superscriptaddress]{revtex4-1}

\usepackage{amsmath}

\usepackage{amsfonts}

\usepackage{amssymb}
\usepackage{color,soul}
\usepackage{graphicx}
\usepackage{empheq}
\usepackage{verbatim}

\usepackage{subfigure}

\newcommand{\vc}{\mathbf}

\allowdisplaybreaks 

\begin{document}

\title{Influence of coupling on thermal forces and dynamic friction in plasmas with multiple ion species}

\author{Grigory~Kagan}
\email[E-mail: ]{kagan@lanl.gov}
\affiliation{Theoretical Division, Los Alamos National Laboratory, Los Alamos, NM 87545}
\author{Scott D. Baalrud}
\affiliation{Department of Physics and Astronomy, University of Iowa, Iowa City, IA 52242}
\author{J{\'e}r{\^o}me Daligault}
\affiliation{Theoretical Division, Los Alamos National Laboratory, Los Alamos, NM 87545}

\date{\today}

\begin{abstract}
The recently proposed effective potential theory [Phys. Rev. Lett. 110, 235001 (2013)] is used to investigate the influence of coupling on inter-ion-species diffusion and momentum exchange in multi-component plasmas. Thermo-diffusion and the thermal force are found to diminish rapidly as strong coupling onsets. For the same coupling parameters, the dynamic friction coefficient is found to tend to unity.
These results provide an impetus for addressing the role of coupling on diffusive processes in inertial confinement fusion experiments. 
\end{abstract}

\maketitle


\section{Introduction}

Transport effects associated with multiple ion species have recently been attracting much attention in inertial confinement fusion (ICF). In particular, it has been suggested that strong background gradients introduced during the implosion can separate fusion fuel constituents, resulting in yield degradation~\cite{amendt2010plasma, amendt2011potential, rygg2006tests, herrmann2009anomalous, casey2012evidence, kagan2012electro, kagan2014thermo, rosenberg2014exploration, hoffman2015approximate, rinderknecht2015ion, hsu2016observation,joshi2017observation,turnbull2017refractive}.  The same \emph{ion diffusion} mechanisms that govern species separation in the fuel underlie  mixing at the shell/fuel interface~\cite{li2002effects, regan2004dependence, ma2013onset, smalyuk2014measurements, albright2006mass, molvig2014nonlinear,vold2015plasma}, as well as at the inner boundary of the hohlraum~\cite{ amendt2014low}. In turn, describing these mix phenomena is often regarded as a crucial challenge faced by the ICF program. 

{The present paper provides a physics insight into how the basic diffusion mechanisms and closely related momentum exchange  can be affected by ion coupling.} Our earlier work on ion diffusion in \emph{weakly coupled} plasmas has found that thermo-diffusivity is comparable to baro-diffusivity in low-Z mixtures and much larger than baro-diffusivity in low-Z/high-Z mixtures~\cite{kagan2014thermo}. This finding came in contrast to what had been known from the conventional theory of neutral gas mixtures, in which thermo-diffusion is usually much less significant than baro-diffusion~\cite{zel2002physics}. Subsequent work~\cite{simakov2016hydrodynamic-1,simakov2016hydrodynamic-2} gave identical prediction~\cite{simakov2006verification,kagan2017comparison}. However, during the course of an implosion, the fuel plasma of hydrogen and helium isotopes can become sufficiently dense that ion correlations can influence transport rates. Interfacial plasmas involve dense high-Z components such as carbon or silicon ions from the ablator, or gold or uranium ions from the hohlraum wall, and are also very likely to be strongly coupled.

 On the one hand, the finite coupling can result in the ideal gas equation-of-state being inadequate for inertially confined plasmas, thus noticeably affecting  the implosion performance~\cite{hu2010strong}. On the other hand, it greatly complicates the physics behind the transport properties as the particle collisions can no longer be considered binary~\cite{ichimaru2004statistical,baus1980statistical}. {To describe evolution of relative species concentrations during the implosion one needs a full set of hydrodynamic equations with appropriate transport fluxes. In turn, developing such a framework requires both thermodynamic and kinetic analyses. } 
 
{Here we present the kinetic analysis of the momentum exchange in a binary ionic mixture with electrons providing neutralizing background. We utilize the recently proposed effective potential theory (EPT)~\cite{baalrud2013effective, daligault2016ionic}, which models many-body correlation effects by treating binary interactions as occurring via the potential of mean force in place of the screened, Debye-H\"uckel potential.
 Our study indicates that with substantial coupling  the thermal force, and therefore thermo-diffusion, rapidly diminishes, making it similar to the conventional case of a neutral gas mixture~\cite{zel2002physics}. These results motivate a need to investigate further the influence of strong coupling on diffusive processes in ICF. }

A physical interpretation for the above finding can be suggested by recalling that thermo-diffusion arrises due to the deviation of a species distribution function away from Maxwellian~\cite{hirschfelder1954molecular, devoto1966transport, ferziger1972mathematical}. Since the collision rate is low in weakly coupled plasmas, the distribution functions are easily perturbed by thermodynamic forces. Consequently, the thermo-diffusion rate is often of comparable importance with the other diffusive processes. In contrast, as strong coupling onsets the Coulomb collision frequency approaches a significant fraction of the plasma frequency~\cite{baalrud2013effective}. The plasma frequency, in turn, defines the fastest timescale for collective particle dynamics. Thus, the plasma becomes so collisional that thermodynamic forces cannot deviate the distributions from Maxwellian substantially, causing thermo-diffusion to diminish.

This interpretation is supported by considering the other diffusion-relevant transport quantity, dynamic friction, which counteracts species separation. In gases, the friction between  two species can usually be evaluated by considering two Maxwellian distributions drifting with respect to each other, whereas in the weakly coupled plasma higher order corrections to Maxwellian are known to give an order unity contribution~\cite{braginskii1965transport}. One consequence of this complication is that all ion species are intertwined: the friction between two given species 
is influenced not only by the inter-species but also by intra-species collisions. Furthermore, in a plasma with three and more ion species, the friction between two given  species would depend not only on these species' thermodynamic states, but also on the states of all other ion species.  In agreement with the logic of the preceding paragraph, this feature is found to disappear for coupling parameter $\Gamma \gtrsim 1$. On the other hand, for such $\Gamma$, the species are intertwined at a deeper level since  the effective interaction potential, and therefore collision frequency, between and within two ion species depend on other species present in the plasma.  The case of small, though finite, $\Gamma$ is thus the most complex from the transport point of view as both of the above mechanisms are generally at play.

In the section to follow we present EPT based evaluation of the inter-ion-species transport in a plasma with two ion species. Then, in Section III, we discuss physical interpretations and practical implications for ICF of the obtained results.

\section{Evaluation of the transport coefficients}

We  consider a plasma with two ion species with the ion charges $Z_1$ and $Z_2$, ion masses $m_1$ and $m_2$, number densities $n_1$ and $n_2$, and mass densities $\rho_1$ and $\rho_2$, where subscripts ``$1$" and ``$2$" denote the light and heavy ion species, respectively. There is only one independent ion concentration, so we will operate with the light species mass fraction $c = \rho_1/\rho$, where $\rho = \rho_1 + \rho_2$ is the total mass density. Similarly, there is only one independent ordinary diffusion coefficient and one independent thermo-diffusion coefficient, so for definitiveness we will be considering $D_{12}$ and $D_{1}^{(T)}$. Notation and transport formulas for the general $N$-ion-species case can be found in literature~\cite{hirschfelder1954molecular, devoto1966transport, ferziger1972mathematical, zhdanov2002transport, kagan2016transport} and, for this paper to be self-contained, are also summarized in the Appendix~\ref{app: formulary} in the   form convenient for practical use.

The leading order approximation to the ordinary diffusion coefficient $[D_{12}]_{1}$ physically corresponds to the situation where the two diffusing species are described by Maxwellian distributions drifting with respect to each other. The mathematical expression can be found in earlier works~\cite{hirschfelder1954molecular, devoto1966transport, ferziger1972mathematical} as well as retrieved from  Eq.~(\ref{eq: ordin-diff}) of the Appendix~\ref{app: formulary} by setting $\xi=1$:
\begin{equation}
\label{eq: ordin-diff-lead}
[D_{12}]_{1} = - \frac{n_i  k_B T_i m_1 (1-c) }{ \rho \mu_{12} \nu_{12}},
\end{equation}
where $n_i = n_1 + n_2$ is the total number density of the ion species, $k_B$ is the Boltzmann constant, and the collision frequency between plasma species $\alpha$ and $\beta$ is defined by
\begin{equation}
\label{eq: nu}
\nu_{\alpha \beta} = \frac{4 \sqrt{2\pi} Z_{\alpha}^2 Z_{\beta}^2 e^4 \gamma_{\alpha\beta}^{3/2} n_{\beta}}
{3   \mu_{\alpha\beta}^2 } 
\Xi_{\alpha \beta}.
\end{equation}
In  Eq.~(\ref{eq: nu}), $\mu_{\alpha\beta} = m_{\alpha} m_{\beta}/(m_{\alpha} + m_{\beta})$ is the reduced mass and $\gamma_{\alpha\beta} = \gamma_{\alpha} \gamma_{\beta}/(\gamma_{\alpha} + \gamma_{\beta})$ with $\gamma_{\alpha} \equiv m_{\alpha} /(k_B T_{\alpha})$  and $T_1 = T_2 \equiv T_i$ need to be set for ion species with comparable masses. Finally, $\Xi_{\alpha \beta} \equiv \Xi_{\alpha \beta}^{(1,1)}$ is the lowest order generalized Coulomb logarithm, which was introduced in Ref.~\cite{baalrud2013effective}. Equation~(\ref{eq: nu}) reduces to the familiar expression in the weakly coupled limit, in which $\Xi_{\alpha \beta}$ becomes the conventional Coulomb logarithm $\ln{\Lambda}$~\cite{baalrud2014extending}. 

In expression~(\ref{eq: ordin-diff-lead}), particle collisions manifest themselves solely through the inter-species collision frequency $\nu_{12}$. The more non-trivial collisional effects appear in higher order approximations to the ordinary diffusion coefficient. These account for the deviation of the species' distribution functions from Maxwellian, which is established through the interplay between the inter- and intra-species collisions. To quantify the resulting correction to the ordinary diffusion coefficient  we introduce 
\begin{equation}
\label{eq: dyn-friction}
A_{12} = [D_{12}]_{1}/D_{12},
\end{equation}
so the full diffusion coefficient can be written as 
\begin{equation}
\label{eq: ordin-diff-true}
D_{12} = - \frac{n_i  k_B T_i m_1 (1-c) }{ A_{12} \rho \mu_{12} \nu_{12}} .
\end{equation}
Thermo-diffusion is conveniently quantified with
\begin{equation}
\label{eq: thermal-force}
B_{1}^{(i)} = \frac{D_{1}^{(T)}}{D_{12}} \frac{c n_i}{n_1}.
\end{equation}

To elucidate the physics behind the dimensionless parameters $A_{12}$ and $B_{1}^{(i)}$, it is useful to write the fluid momentum conservation equations for each ion species. Then it can be shown that in order to recover the  thermodynamic expression for the diffusive flux~\cite{Note1}, the rate of the collisional momentum exchange between the two ion species must take the form~\cite{kagan2012electro}  
\begin{equation}
\label{eq: friction-bin}
\vec{ R}_{12} =-A_{12}\mu_{12} n_{1} \nu_{12} ( \vec{u}_{1} -  \vec{u}_{2}) - 
B_{1}^{(i)} n_{1} k_B \nabla T_{i},
\end{equation}
where the first and second terms on the right side are referred to as the dynamic friction and the ion-ion thermal force, respectively, and $ \vec{u}_{\alpha} $ stands for the fluid velocity of species $\alpha$. Eqs.~(\ref{eq: dyn-friction}) and (\ref{eq: thermal-force}) on the one hand and Eq.~(\ref{eq: friction-bin}) on the other show the close connection between the momentum exchange and diffusion. This will be used for both developing physics understanding of the basic trends in the diffusion coefficients and quantifying these trends through the dimensionless parameters $A_{12}$ and $B_{1}^{(i)}$.

We notice here that a recent study applying EPT to ordinary diffusion~\cite{beznogov2014effective} concluded that the higher order corrections to $D_{12}$ are small, which in our  terms would mean that $A_{12}$ is identically equal to one. Likely, this is because this reference concentrated on the strongly coupled regime only, thus not considering the physics of the transition between the weakly and strongly coupled regimes.

We now proceed to evaluation of $A_{12}$ and $B_{1}^{(i)}$ using EPT.
The generalized Coulomb logarithms are computed as described in \cite{baalrud2013effective}:
\begin{equation}
\Xi_{\alpha \beta}^{(l,k)} = \frac{1}{2} \int_0^\infty d\xi\, \xi^{2k+3} e^{-\xi^2} \bar{\sigma}_{\alpha \beta}^{(l)}/\sigma_{o}  \label{eq:xi}
\end{equation}
where
\begin{equation}
\bar{\sigma}_{\alpha \beta}^{(l)} = 2\pi \int_0^\infty db\, b[1 - \cos^l (\pi - 2 \Theta)]
\end{equation}
is the momentum transfer cross section, and 
\begin{equation}
\Theta = b \int_{r_o}^\infty dr\, r^{-2} [1 - b^2/r^2 - 2\phi_{\alpha \beta}(r)/(m_{\alpha \beta} u^2)]^{-1/2} \label{eq:theta}
\end{equation}
is the scattering angle. Here, $\sigma_o = \pi Z_\alpha^2 Z_\beta^2 e^4 \gamma_{\alpha \beta}^2/(4 \mu_{\alpha \beta}^2)$ is a defined reference cross section. 

The input to the theory is the interaction potential, $\phi_{\alpha \beta}$ in Eq.~(\ref{eq:theta}), which is taken to be the potential of mean force. The potential of mean force is obtained by taking two particles at fixed positions and averaging over the positions of all other particles in the system. At equilibrium, it is related to the radial distribution function as $g_{\alpha \beta}(r) = \exp(-\phi_{\alpha \beta}(r)/k_BT)$. 
Here, we use the hypernetted chain (HNC) approximation to model the radial distribution function \cite{hansen1976}:
\begin{subequations}
\label{eq:hnc}
\begin{eqnarray}
g_{\alpha \beta}(\vec{r})&=&\exp [- v_{\alpha \beta}(\vec{r})/k_\textrm{B} T + h_{\alpha \beta}(\vec{r}) - c_{\alpha \beta}(\vec{r}) ]\label{eq:hncgr}\\
\hat{h}_{\alpha \beta}(\vec{k})&=&\hat{c}_{\alpha \beta}(\vec{k}) - \sum_j n_j \hat{h}_{\alpha j}(\vec{k}) \hat{c}_{j \beta} (\vec{k}) \,,  \label{eq:hnchk} 
\end{eqnarray}
\end{subequations}
where $h_{\alpha \beta}(r)=g_{\alpha \beta}(r)-1$ and $\hat{h}_{\alpha \beta} (\vec{k})$ denotes the Fourier transform of $h_{\alpha \beta}(\vec{r})$.

Once the generalized Coulomb logarithms are calculated 
Eqs.~(\ref{eq: ordin-diff})-(\ref{eq: thermo-diff}) along with Eqs.~(\ref{eq: M-elem-1})-(\ref{eq: bracket-1-end}) are employed to recover the coefficients of interest.
For a mixture of a given set of ion species, they generally depend on the relative species concentrations and the Coulomb coupling parameter
\begin{equation}
\label{eq: Gamma}
\Gamma = \frac{e^2}{k_B T_i a},
\end{equation}
where  $a \equiv (4\pi n_i/3)^{-1/3}$. It should be noted that this definition does not include the charge numbers, so in the one component plasma (OCP) the standard coupling parameter $\Gamma_{\alpha}$ is recovered through $\Gamma_{\alpha} = Z_{\alpha}^2 \Gamma$. Also, for the purpose of demonstration, the ion species  are assumed fully ionized. Finally, unless otherwise specified, the transport coefficients are calculated in the third order Chapman-Enskog approximation ($\xi=3$) by utilizing the matrix elements from Appendix~\ref{app: matrix-elem}. 

{It should be noted that  the transport coefficients being discussed here have, in general, a non-unity thermodynamic prefactor~\cite{hansen1985} that we do not include. Recent refinement of the EPT theory based on Enskog's equation~\cite{baalrud2015modified} can be applied to model these terms using   the expressions from Ref.~\cite{lopez1983}, but resulting changes in transport coefficients are insignificant over the range of coupling strengths being discussed here. 
We also notice that the diffusion coefficients $D_{12} $  and $D_{1}^{(T)}$ alone are not sufficient for quantitative modeling of multi-component plasmas. Complete expression for the diffusive flux involves other driving terms, which are subject to thermodynamic rather than kinetic calculation. This expression along with the equation for  evolving $c$ are given in the Appendix~\ref{app: concentration-evolution}}.

Depending on the plasma composition, different new features can appear in  the dynamic friction and thermal forces as compared to their weakly coupled counterparts. To illuminate the most essential  trends, we now discuss $A_{12}$ and $B_{1}^{(i)}$  for three representative binary ionic mixtures (BIMs): DT, D\textsuperscript{3}He and DKr. 

\subsection*{Isotopic mixture}

We first consider DT, which is the most common choice for the fusion fuel.{In cryogenic ICF implosions the DT fuel can be weakly coupled as well as strongly coupled with $\Gamma \sim 6$~\cite{hu2011first}}.
In isotopic mixtures, all the effective interaction potentials between and within any ion species are equal and the generalized Coulomb logarithms 
depend on $\Gamma$ only, $\Xi_{\alpha\beta}^{(l,k)} = \Xi_{\alpha\beta}^{(l,k)}(\Gamma)$. For the DT case it is therefore given by the same function as in the one-component plasma model for hydrogen considered in our earlier work~\cite{baalrud2013effective}.
\begin{figure}[h!]
\begin{center}
\subfigure[]{
\includegraphics[scale=0.6]{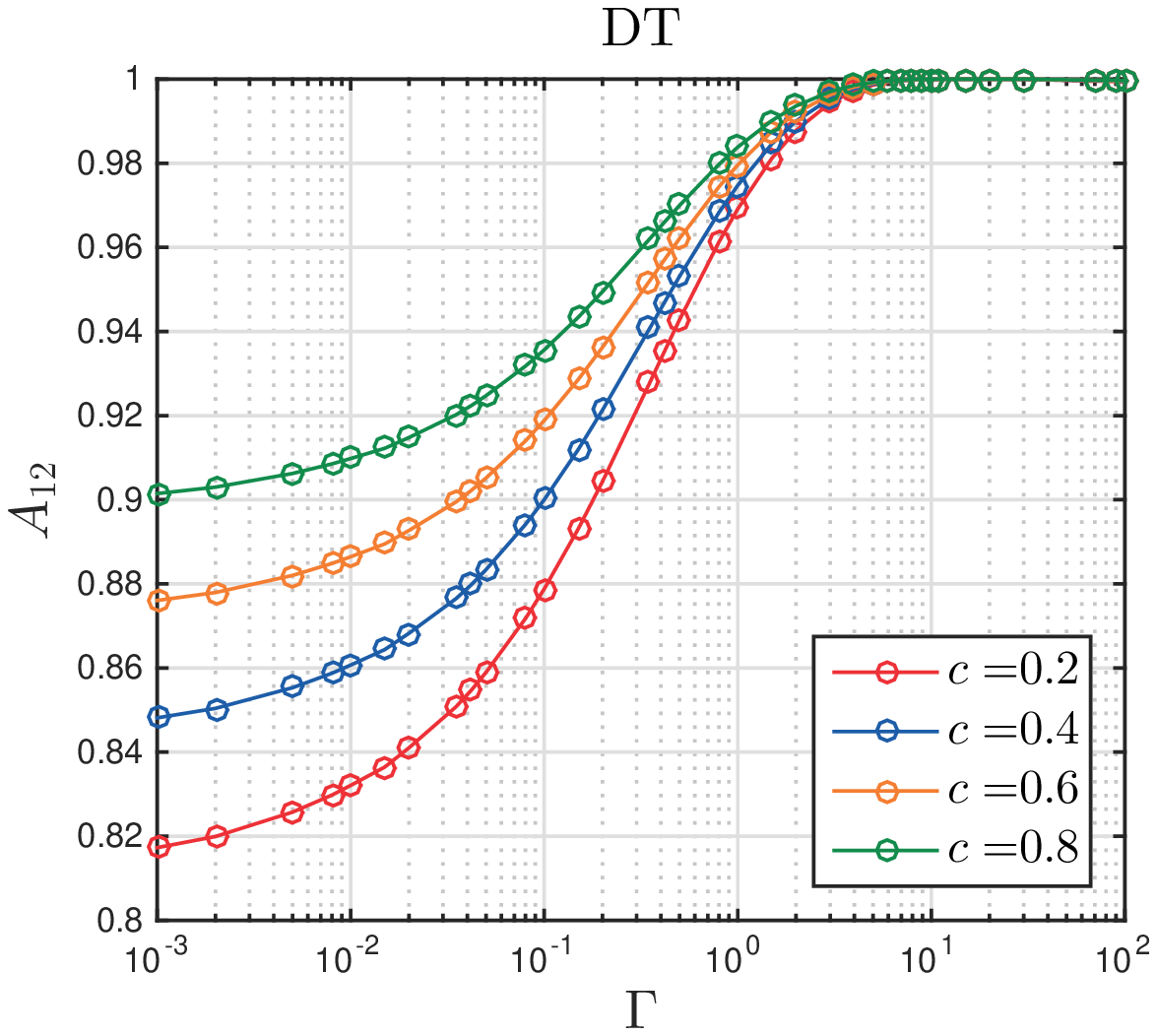}}
\subfigure[]{
\includegraphics[scale=0.6]{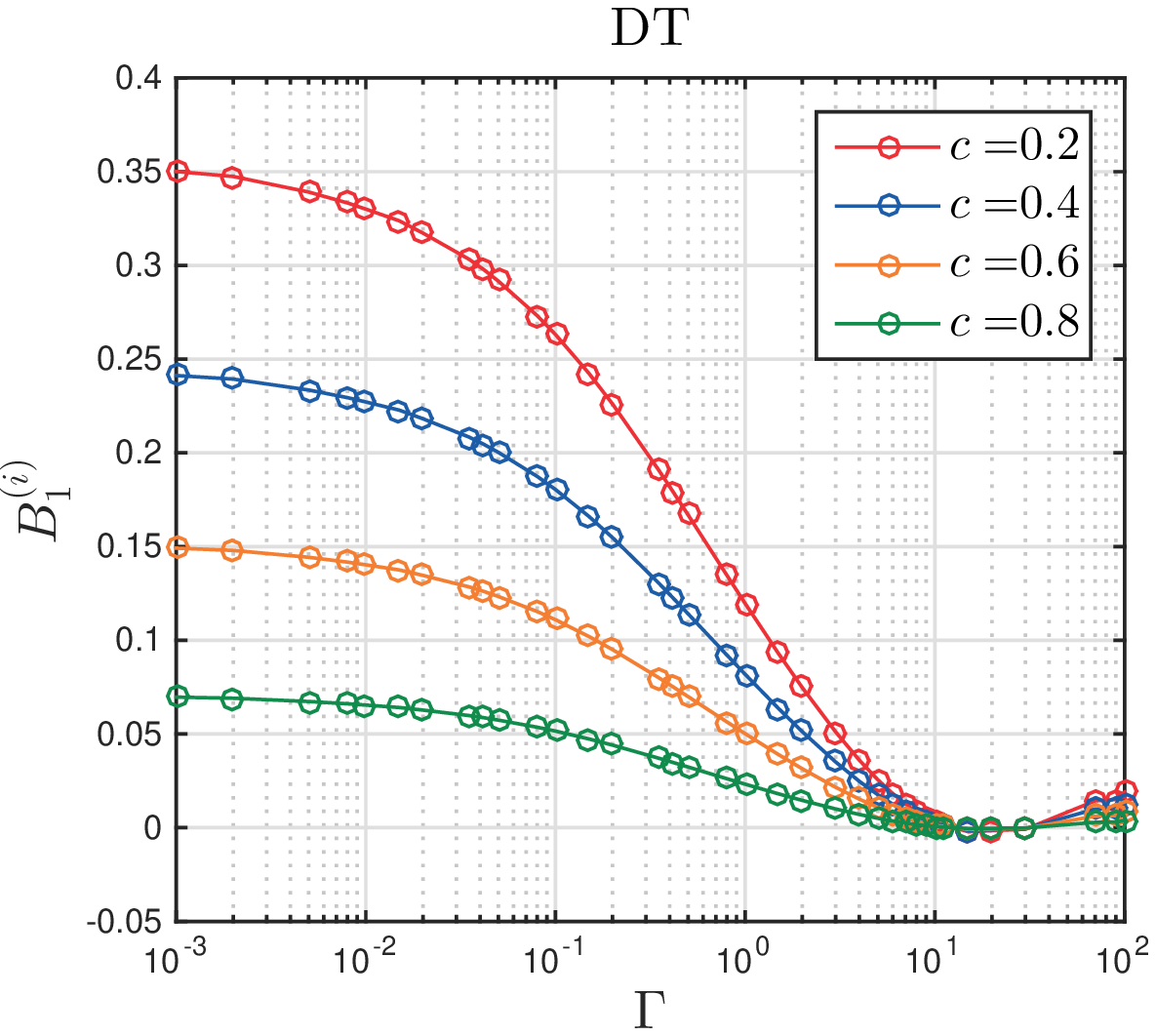}}
\caption{\label{fig: DT_friction_and_thermal} 
Dynamic friction coefficient $A_{12}$ (a) and thermal force coefficient $B_{1}^{(i)}$ (b) for the DT mixture as functions of $\Gamma$ for several values of $c$. }
\end{center}
\end{figure}
\label{app: formulary-diff}

The resulting dynamic friction and the ion-ion thermal force coefficients are shown in Fig.~{\ref{fig: DT_friction_and_thermal}. It should first be observed that, as $\Gamma$ approaches 10, $A_{12}$ tends to unity, while $B_{1}^{(i)}$ rapidly diminishes. Hence, with substantial coupling the intuitive expression for the dynamic friction between two ion species 
$\vec{R}^{(u)}_{12} = \mu_{12} \nu_{12} n_{1} (\vec{u}_{2} - \vec{u}_{1})$ 
becomes precise. From the diffusion perspective, it means that the ordinary diffusion coefficient $D_{12}$  can  be obtained readily from Eq.~({\ref{eq: ordin-diff-lead}) once the inter-species collision frequency $ \nu_{12}$ is known. 

As discussed earlier, an $A_{12}(c)$ differing from unity arrises from higher order corrections to the distribution function. The observed trend in $A_{12}$ is thus indicative of the role of these corrections being diminished by the ion-ion correlations. This interpretation is supported by the fact that the $A_{12}$ saturation to unity takes place at the same coupling parameters as vanishing of the thermal force, which is an inherently higher order transport phenomenon~\cite{hirschfelder1954molecular, devoto1966transport, ferziger1972mathematical}. Although according to Fig.~{\ref{fig: DT_friction_and_thermal} the absolute value of $B_{1}^{(i)}$ starts growing after reaching $0$, it remains much smaller than in the weakly coupled case. Moreover, this trend is only observed for $\Gamma \gtrsim 30$ where the effective potential theory becomes invalid~\cite{baalrud2013effective}.

\subsection*{Weakly asymmetric mixture with different ion charge numbers}

When the charge numbers of the ion components differ, the effective interaction potential depends on the concentration as well as $\Gamma$, 
 i.e. $\Xi_{\alpha\beta}^{(l,k)} = \Xi_{\alpha\beta}^{(l,k)}(\Gamma,c)$ and are also different for different $\alpha$ and $\beta$. In the intermediate case when higher order corrections to the species' distribution functions are still important while coupling is no longer negligible, all these Coulomb logarithms should generally contribute to the ion transport; e.g. for evaluating inter-ion-species transport in a binary mixture of species ``1" and ``2" one needs to know not only $\Xi_{12}^{(l,k)}$, but also $\Xi_{11}^{(l,k)}$ and $\Xi_{22}^{(l,k)}$.

A weakly  asymmetric example of such a mixture commonly used in ICF experiments is  D\textsuperscript{3}He. Fig.~\ref{fig: coulomb_logs_D3He} presents the lowest order Coulomb logarithms for this case. 
\begin{figure}[h!]
\begin{center}
\includegraphics[scale=0.6]{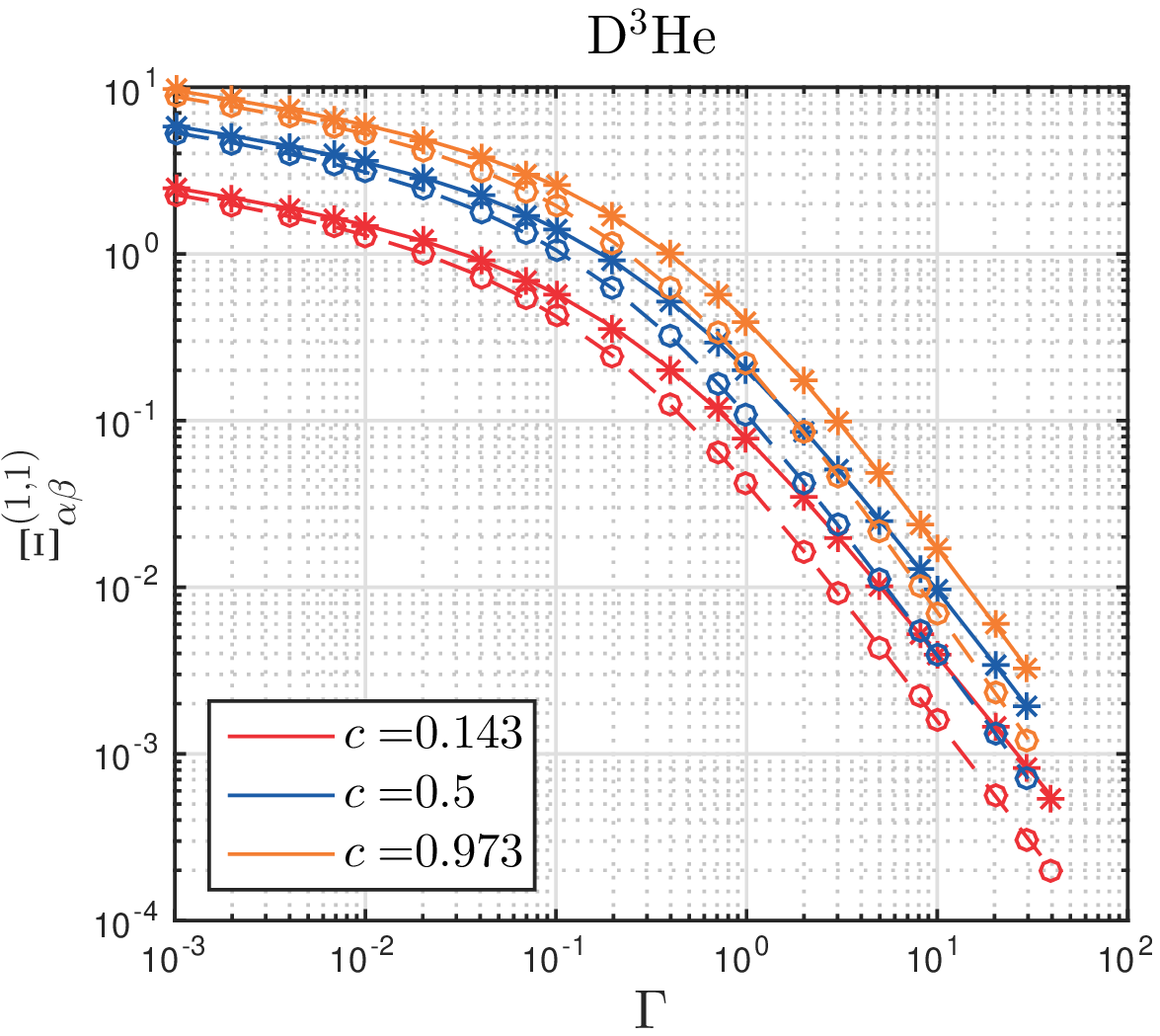}
\caption{ \label{fig: coulomb_logs_D3He}   Lowest order generalized Coulomb logarithms $\Xi_{11}^{(1,1)}$ (asterisks, solid line) and 
$\Xi_{12}^{(1,1)}$ (circles, dashed line) in the D\textsuperscript{3}He mixture as functions of $\Gamma$ for several values of $c$. }
\end{center}
\end{figure}
It can be observed that $\Xi_{\alpha\beta}^{(1,1)}$ has an order unity dependence on the concentration throughout the entire $\Gamma$ range; the Coulomb logarithms for mixtures with a trace amount of \textsuperscript{3}He ($c = 0.937$) are about 4-5 times larger than for mixtures with a trace amount of D ($c = 0.143$). 
 For a fixed $\Gamma$, i.e. fixed total ion density, the system consisting of just \textsuperscript{3}He demonstrates stronger correlation effects than the system consisting of just D since the OCP coupling scales as $Z^2$.
It is also worth noticing that 
$\Xi_{11}^{(1,1)}$ and $\Xi_{12}^{(1,1)}$ coalesce in the weakly coupled limit, in which the conventional Coulomb logarithm can be used for all types of collisions.

According to Fig.~\ref{fig: D3He_friction_and_thermal}, the dynamic friction and thermal force coefficients show the same general trends as recovered earlier for the DT mixture: $A_{12}$ tends to 1 and $B_{1}^{(i)}$ vanishes, respectively, as the coupling parameter becomes of order unity. 
As in the DT case, it means that with order unity coupling the classical diffusion coefficient no longer has a non-trivial dependence on the concentration when expressed in terms of the inter-species collision frequency. However, \emph{unlike} the case of the species with equal charge numbers, the non-trivial dependence on the concentration now appears in the collision frequency through the Coulomb logarithm $\Xi_{12}^{(1,1)}$ as indicated by Fig.~\ref{fig: coulomb_logs_D3He}. \begin{figure}[h!]
\begin{center}
\subfigure[]{
\includegraphics[scale=0.6]{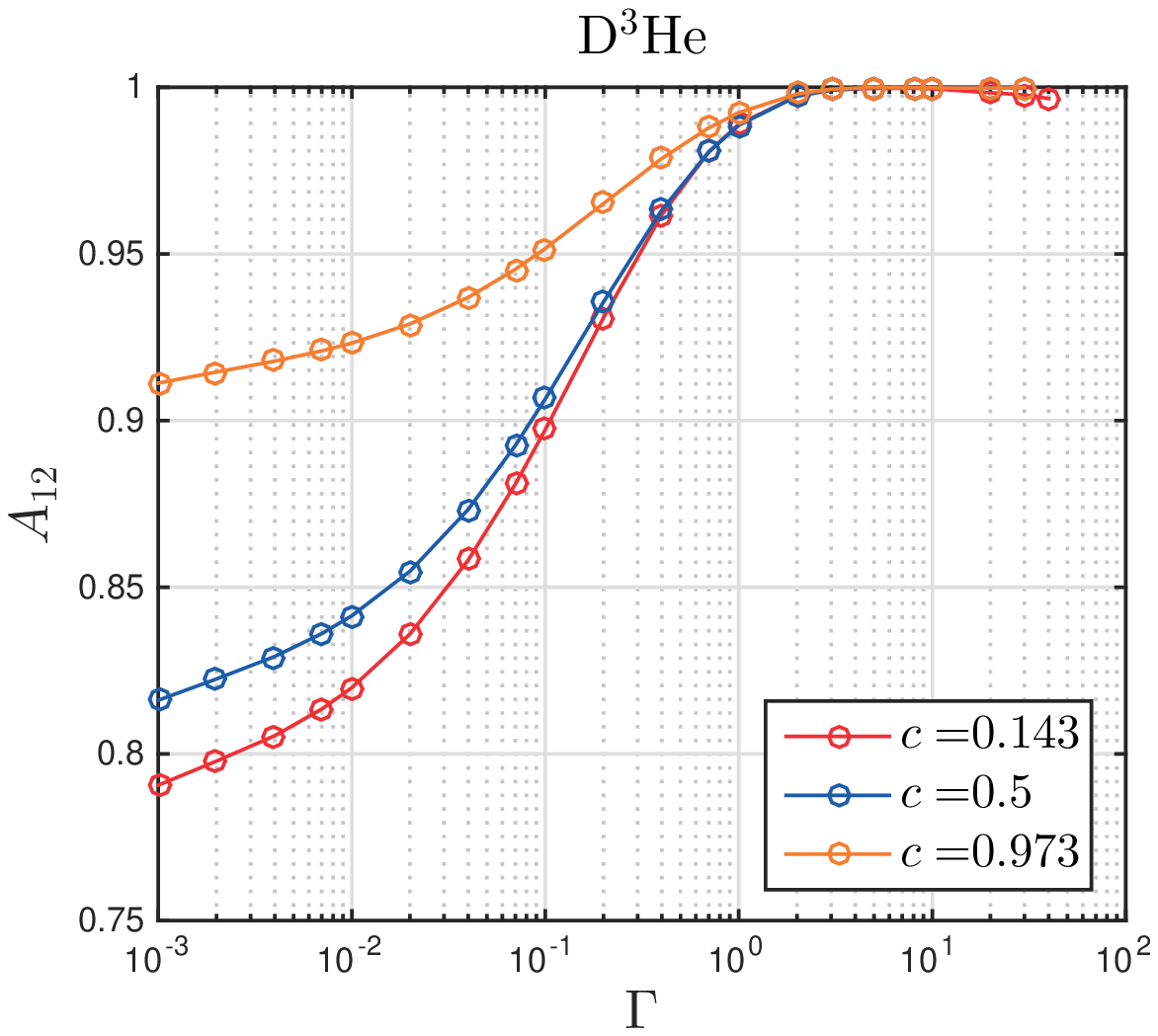}}
\subfigure[]{
\includegraphics[scale=0.6]{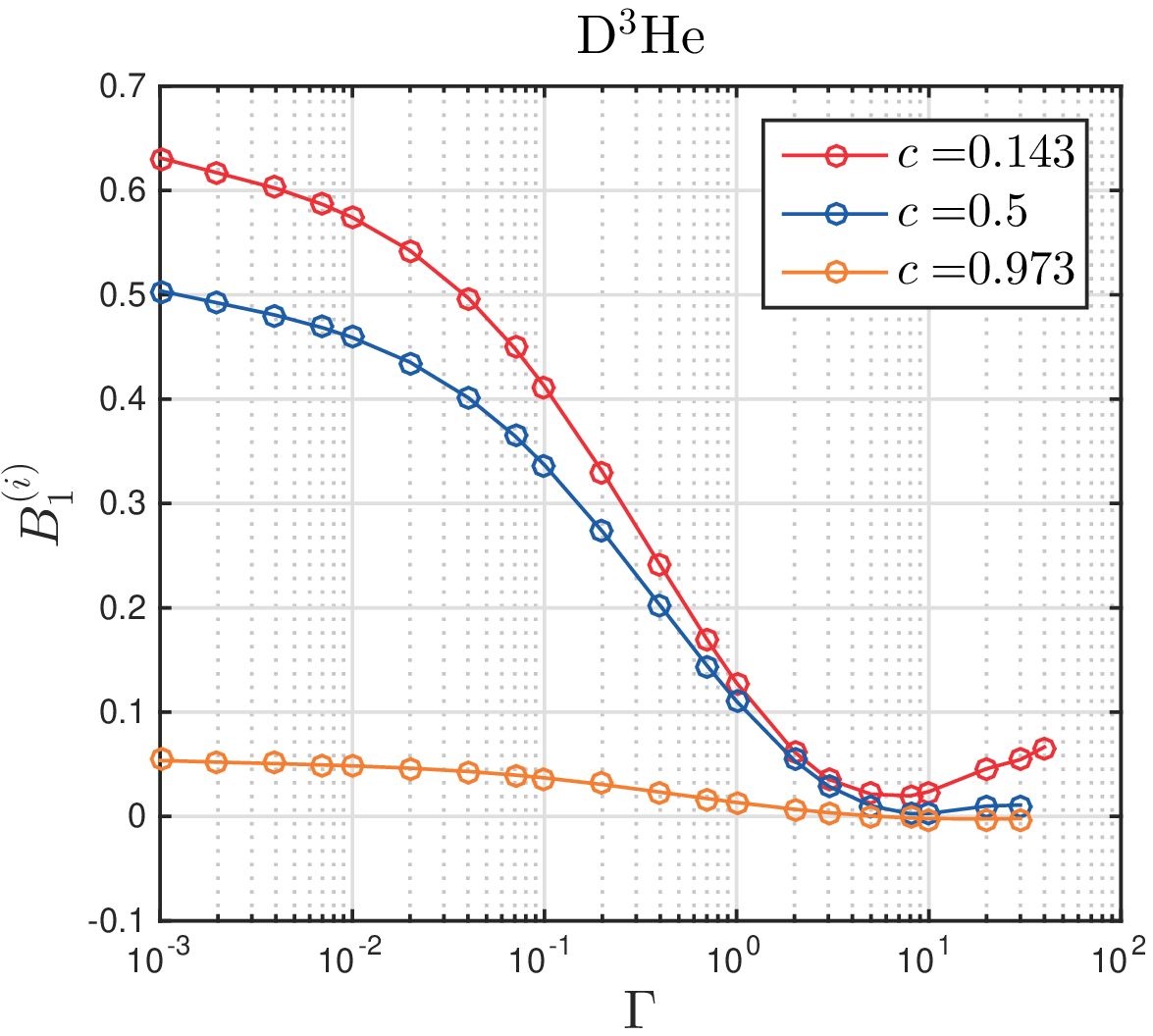}}
\caption{\label{fig: D3He_friction_and_thermal} 
Dynamic friction coefficient $A_{12}$ (a) and thermal force coefficient $B_{1}^{(i)}$ (b) for the D\textsuperscript{3}He mixture as functions of $\Gamma$ for several values of $c$.}
\end{center}
\end{figure}

\subsection*{Strongly asymmetric mixture}
Finally, to gain an insight into diffusion at the low-Z/high-Z interfaces we consider the mixture of species with largely disparate masses and charge numbers such as DKr,  in which the concentration dependence is expected to be even more pronounced. We first plot the lowest order generalized Coulomb logarithms in Fig.~\ref{fig: coulomb_logs_DKr}. It should  be observed that we are able to evaluate  $\Xi_{\alpha\beta}^{(l,k)}$ up to $\Gamma \sim 0.5 \times  10^{-1}$ only, i.e. a coupling strength of Kr $Z^2 \Gamma \approx 60$, beyond which the EPT is not expected to be accurate. 

One interesting feature is that when the Kr becomes strongly coupled it has a significant influence on the DD effective potential, leading to a flattening of the DD generalized Coulomb logarithm $\Xi_{\textrm{11}}^{(1,1)}$. This behavior was not observed for lower-Z mixtures. It is also not observed for the case where Kr number fraction is only about $2 \times 10^{-4}$  (D mass fraction $c=0.99$), indicating an impurity limit is reached in which the small concentration of Kr does not significantly influence the DD interaction. 
\begin{figure}[h!]
\begin{center}
\includegraphics[scale=0.6]{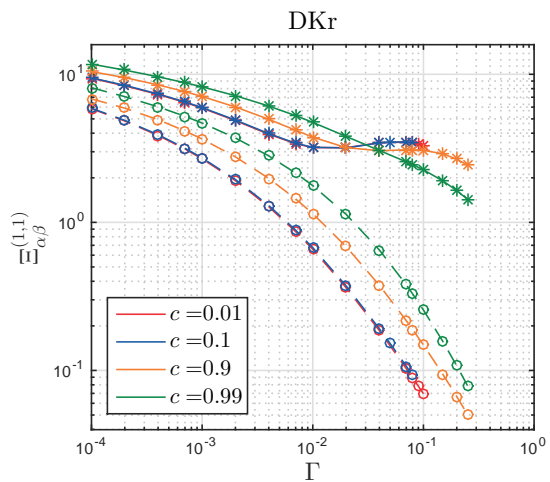}
\caption{ \label{fig: coulomb_logs_DKr}   Lowest order generalized Coulomb logarithms $\Xi_{11}^{(1,1)}$ (asterisks, solid line) and 
$\Xi_{12}^{(1,1)}$ (circles, dashed line) in the DKr mixture as functions of $\Gamma$ for several values of $c$. }
\end{center}
\end{figure}

The dynamic friction and thermal force coefficients for the DKr mixture are shown in Fig.~\ref{fig: DKr_friction_and_thermal}.  One can see that the results for $c=0.01$ and $c=0.1$ coalesce and are also rather close to the results for $c = 0.9$ (except for the maximum coupling being higher for larger $c$, which is explained earlier in this subsection). The reason being that, due to the large charge number of Kr, D-Kr collisions dominate over D-D  up to quite high mass fractions of D, making the D ions behave as test particles in the swamp of Kr ions.
\begin{figure}[h!]
\begin{center}
\subfigure[]{
\includegraphics[scale=0.6]{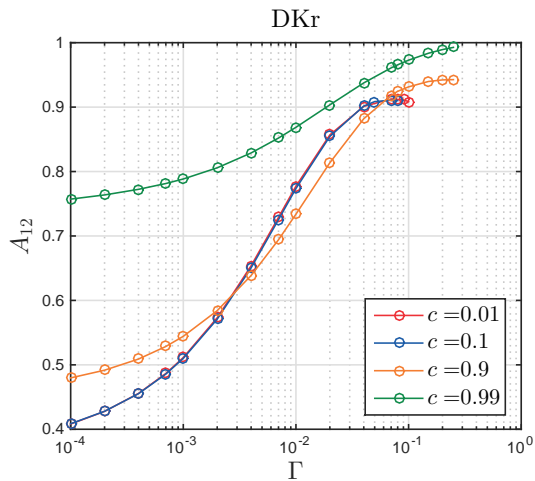}}
\subfigure[]{
\includegraphics[scale=0.6]{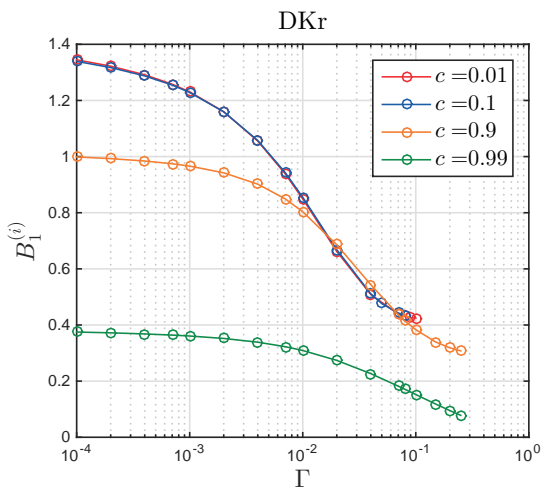}}
\caption{\label{fig: DKr_friction_and_thermal} 
Dynamic friction coefficient $A_{12}$ (a) and thermal force coefficient $B_{1}^{(i)}$ (b) for the DKr mixture as functions of $\Gamma$ for several values of $c$.}
\end{center}
\end{figure}

Plots in Fig.~\ref{fig: DKr_friction_and_thermal} are indicative of the same trends as were earlier recovered for the DT and D\textsuperscript{3}He mixtures.{However, unlike their weakly asymmetric counterparts, $A_{12}$ and $B_{1}^{(i)}$ do not saturate fully to $1$ and $0$, respectively, over the $\Gamma$ range accessible with EPT. To confirm that the qualitative results found here for the weakly asymmetric mixtures remain intact in the strongly asymmetric case, one would need to use a different approach such as molecular dynamics simulations \cite{daligault2012diffusion, whitley2015localization, haxhimali2015shear, ticknor2016transport}.}

\subsection*{Dynamic friction and thermal force coefficients obtained with other formalisms}

The above results are based on the effective binary interaction potential equal to the potential of mean force, as obtained through the HNC closure. One can also use older theories for the effective potential such as that by Debye-H{\"u}ckel~\cite{liboff1959transport} and Paquette~\cite{paquette1986diffusion}. In Fig.~\ref{fig: comparison} we compare $A_{12}$ and $B_{1}^{(i)}$ for the 50:50 DT mixture obtained from the EPT, as described at the beginning of this section, with their counterparts obtained from these other formalisms. Generalized Coulomb logarithms are evaluated through gas-kinetic cross-sections as described in Ref.~\cite{baalrud2013effective} and the transport coefficients are computed with formulas of the Appendix~\ref{app: formulary}. Also shown is the weakly coupled limit $\Gamma \to 0$ considered in our earlier work~\cite{kagan2014thermo}.
\begin{figure}[h!]
\begin{center}
\subfigure[]{
\includegraphics[scale=0.605]{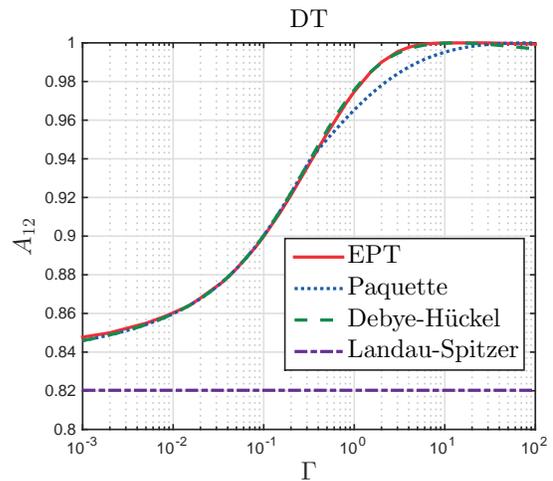}}
\subfigure[]{
\includegraphics[scale=0.6]{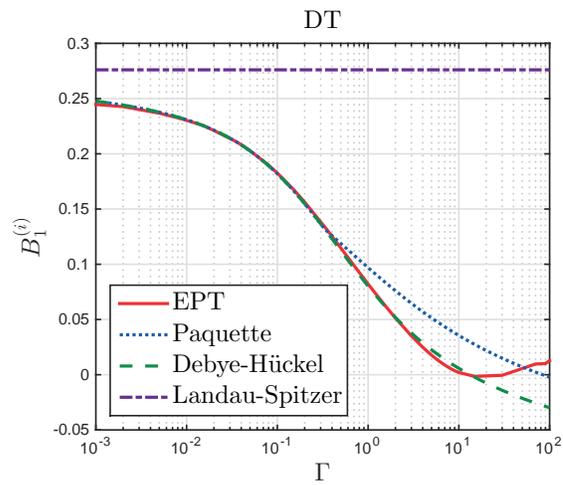}}
\caption{\label{fig: comparison} 
Dynamic friction coefficient $A_{12}$ (a) and thermal force coefficient $B_{1}^{(i)}$ (b) for the 50:50 DT mixture as functions of $\Gamma$ obtained from different microscopic theories.}
\end{center}
\end{figure}

We see that the main trends in the dynamic friction and thermal force coefficients are in excellent agreement further supporting robustness of our conclusions. It should be noticed from Eq.~(\ref{eq: friction-bin}), however, that collisional physics enters the momentum exchange not only through the dimensionless parameters  $A_{12}$ and $B_{1}^{(i)}$ but also through the collision frequency $\nu_{12}$. In turn, this frequency involves  the generalized Coulomb logarithm $\Xi_{12}$. As shown in our earlier work~\cite{baalrud2015effective}, $\Xi_{12}$ obtained from the effective potential theory~\cite{baalrud2013effective} give better agreement with the molecular dynamics simulations than that from the Debye-H{\"u}ckel~\cite{liboff1959transport} and Paquette~\cite{paquette1986diffusion} formalisms. Hence, when evaluating  physical observables, such as the momentum exchange or  diffusive flux, the effective potential theory used in the present paper is preferable.

\section{Discussion}

To understand the physics behind the dynamic friction coefficient becoming 1 for strongly coupled plasmas, we first discuss the physics that makes it less than 1 in the conventional plasmas. For this purpose we  follow and adapt the arguments given by Braginskii to interpret the electron-ion friction in a simple plasma~\cite{braginskii1965transport}. Braginskii first considered Maxwellian electrons drifting with respect to ions at the speed  $\vec{u} \equiv \vec{u}_{e} -  \vec{u}_{i}$  and defined the effective electron collision time $\tau_e$ so that the electron-ion friction would write in a simple form without numerical factors, namely
\begin{equation}
\label{eq: friction-Braginskii}
\vec{ R}_{u} = - A_{ei} m_{e} n_{e} \vec{u} /\tau_e
\end{equation}
with $A_{ei} = 1$.  As rigorous calculation showed, however, a numerical factor does need to be included: $A_{ei} = 0.51$ for $Z=1$, $0.44$ for $Z=2$ ... 0.29 for $Z=\infty$. This was explained by the Coulomb collision frequency scaling inversely with the cube of the particle velocity. When external force (e.g. the electric field) is applied to separate the species,  the faster electrons gain a larger velocity shift than the slower. Consequently, the electron distribution is perturbed so that the faster electrons play a larger role on the average velocity $\vec{u}$, making the friction coefficient smaller than for the uniformly shifted perfect Maxwellian. 

This qualitative picture applies to the friction between two weakly coupled ion species as well. Unlike the electron and ion densities in a simple plasma, which are constrained by quasi-neutrality, the ratio of the ion species densities in binary ionic mixtures can take an arbitrary value. This is why the electron-ion friction in a simple plasma is defined by a single numerical factor that depends on the charge number of the single ion species only, whereas a function of the relative species concentration $A_{12}(c)$ is needed to describe the friction between the ion species in a BIM. But, this function is still in agreement with the physical explanation above in that $A_{12}(c) \leq 1$  for all $c$ as it is demonstrated by Fig. 1  of Ref.~\cite{kagan2014thermo}. 

For a BIM with disparate ion masses, such as DKr, further insight from Braginskii's consideration can be derived. There, it is modifications to the light species distribution function that govern the dynamic friction, because velocities of the heavy ions are much lower, making the details of their distribution unimportant. In turn, when concentration of the heavy component is small, the distribution of the light species is close to Maxwellian since collisions within it dominate over collisions with the heavy ions. As a result, $A_{12}(c)$ should approach $1$ together with $c$, as it is indeed recovered for the DKr case in Fig. 1  of Ref.~\cite{kagan2014thermo}. For mixtures with closer ion masses, such as DT, modifications to both distribution functions contribute to $A_{12}$ and, at least for one of the species, this modification is substantial for any given $c$. Consequently, for weakly asymmetric weakly coupled ionic mixtures $A_{12} < 1$ for all $c$. 

So, while $A_{12}$ may become 1 even in the weakly coupled limit, it is only possible at a certain concentration in a largely asymmetric mixture. Furthermore, Braginskii's explanation seems to suggest that $A_{12}$ should be different from 1 as long as there is a non-trivial dependence of the collision frequency  on the particle velocity, which remains true for the collision frequency~({\ref{eq: ordin-diff-lead}) in coupled plasmas. The seeming contradiction between the newly obtained result and the Braginskii's qualitative theory may be resolved by comparing the plasma and collision frequencies for various coupling parameters.

We present such a comparison for the one-component plasma case in Fig.~\ref{fig: nu-vs-omega}a, where $\omega_{p\alpha} = (4\pi n_{\alpha} Z_{\alpha}^2 e^2/m_{\alpha})^{1/2}$ and $Z_{\alpha} = 1$ for simplicity. To make  correspondence with the transport properties we also present in Fig.~\ref{fig: nu-vs-omega}b the ratio of the third to the first order approximations for the ion viscosity $\eta$   and  heat conductivity  $\lambda$. We see that, as the plasma undergoes transition between the weakly coupled and strongly coupled regimes, 
the collision frequency becomes three orders of magnitude closer to the plasma frequency to saturate at $\sim 10^{-1} \omega_{p\alpha}$ for $\Gamma \gtrsim 1$. For the same $\Gamma$ the higher order approximations for transport coefficients become unnecessary and therefore corrections to the species distribution function -- unimportant. The plasma frequency defines the fastest time scale for collective phenomena in a plasma and  perturbations of the distribution function capable of contributing to transport are slower. One could then hypothesize that with $\nu_{\alpha\alpha}$  reaching the $10^{-1} \omega_{p\alpha}$ ballpark such perturbations are mitigated by the collisions and plasma, once Maxwellized, remains Maxwellian.

We also notice that the saturation observed in Fig.~\ref{fig: nu-vs-omega}a may be explained by the same physics: assuming that the effective binary collision frequency $\nu_{\alpha\alpha}$ is a proper description for the actual, many-body collisions in coupled plasmas, one should expect that $\nu_{\alpha\alpha}$ cannot exceed a certain fraction of $\omega_{p\alpha}$ since these many-body collisions are also a collective plasma process.
\begin{figure}[h!]
\begin{center}
\subfigure[]{
\includegraphics[scale=0.57]{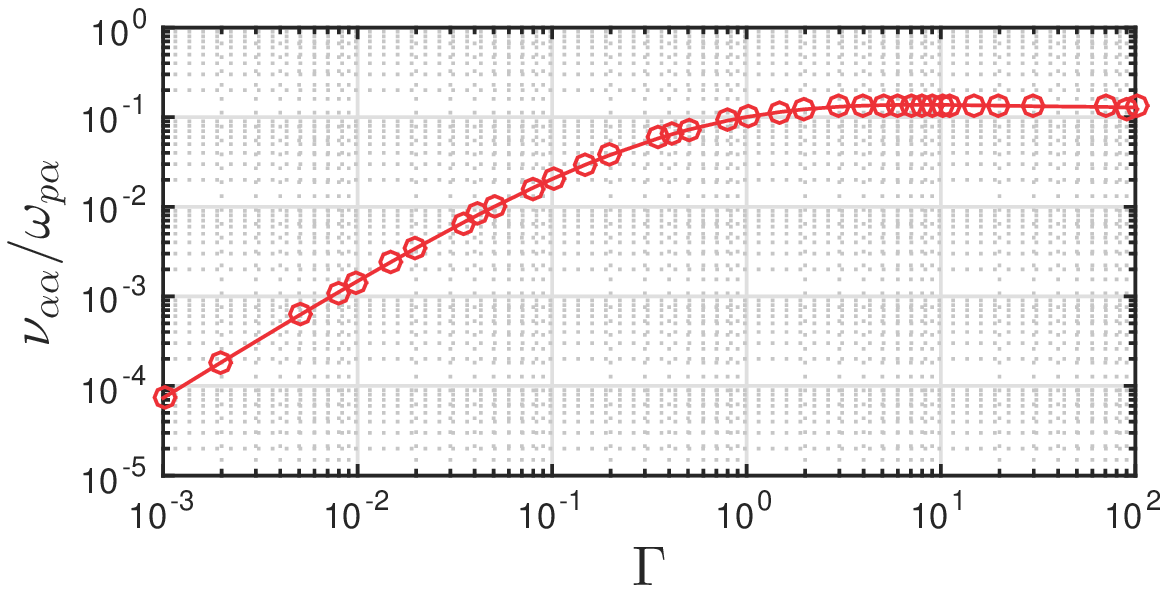}}
\subfigure[]{
\includegraphics[scale=0.57]{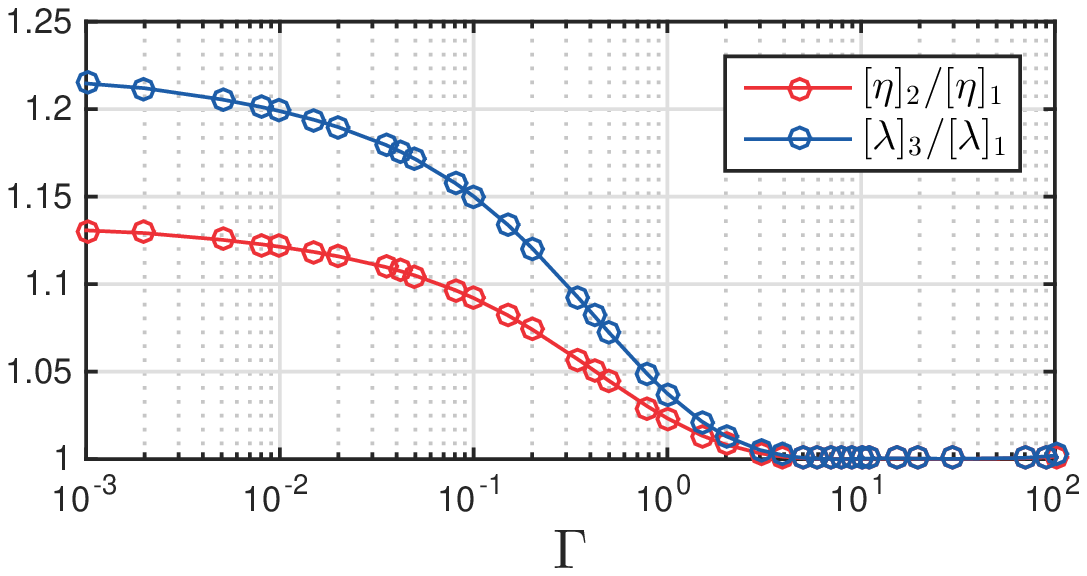}}
\caption{\label{fig: nu-vs-omega} 
Collision vs. plasma frequencies (a) and higher order approximation vs. leading order result for the viscosity  and heat conductivity (b) as functions of $\Gamma$ for the  hydrogen OCP.}
\end{center}
\end{figure}

While one can look for alternative interpretations for the observed trend in $A_{12}$, the very fact that within the EPT framework the role of the higher order corrections is diminished with coupling has been verified by calculating a number of different transport coefficients~\cite{baalrud2014extending,daligault2014viscosity}. These include the OCP viscosity and heat conductivity shown in Fig.~\ref{fig: nu-vs-omega}b and, in particular, thermal force coefficient in a binary ionic mixture presented earlier in this article. Thermal force and closely related thermo-diffusion are known to be  higher order transport phenomena~\cite{hirschfelder1954molecular, devoto1966transport, ferziger1972mathematical}. With the higher order corrections becoming insignificant one would thus expect thermal force and thermo-diffusion to vanish, as has been indeed recovered in Figs.~\ref{fig: DT_friction_and_thermal}b and \ref{fig: D3He_friction_and_thermal}b showing the coefficient $B_{1}^{(i)}$ for the DT and D\textsuperscript{3}He mixtures, respectively. For the DKr mixture, this trend has not be recovered fully. But this rather reflects a limitation of the EPT framework, preventing us from accessing the same coupling parameters for low-Z/high-Z mixtures, for which the saturation has been observed in low-Z mixtures.

The plasma diffusion can be driven by the gradients of the pressure, electrostatic potential and the electron and ion temperatures, as well as by the concentration gradient which  acts to relax the concentration perturbation. In neutral gases, thermo-diffusivity is  known to usually be substantially smaller than baro-diffusivity~\cite{zel2002physics}. On the other hand, in our earlier work we demonstrated that in weakly coupled plasmas the former is comparable to, or even much larger than, the latter~\cite{kagan2014thermo}. The present analysis suggests that in the strongly coupled plasmas the comparison may resort back to the familiar case of the neutral gas mixture diffusion. Assuming that spatial scales associated with the pressure and temperature profiles are similar, the newly presented result likely means that in strongly coupled inertially confined plasmas it is baro-diffusion, rather than thermo-diffusion, that would underlie diffusion-sensitive experimental observations.

\acknowledgements
GK acknowledges many useful discussions with A.A. Stepanenko and V.M. Zhdanov of MEPhI and N.M. Hoffman of LANL. The work of GK and JD was performed under the auspices of the U.S. Dept. of Energy by the Los Alamos National Security, LLC, Los Alamos National Laboratory under Contract No. DE-AC52-06NA25396.
Authors would also like to thank the anonymous referee for suggesting the comparison with other theories; this paper is focused on coupling effects so comparison between our earlier, weakly coupled study~\cite{kagan2014thermo} and the subsequent weakly coupled study~\cite{simakov2016hydrodynamic-1,simakov2016hydrodynamic-2} appeared as a separate Note~\cite{kagan2017comparison}.
 The work of GK was partially supported by the ASC Thermonuclear Burn Initiative and the work of JD was partially supported by the LDRD Grant No. 20150520ER. The work of SDB was supported by the U.S. Department of Energy, Office of Science, Office of Fusion Energy Sciences under Award No. {DE-SC0016159}. 

\appendix \section{Explicit expressions for the diffusion coefficients in terms of the generalized Coulomb logarithms}
\label{app: formulary}

This Appendix provides explicit expressions for the diffusion coefficients in terms of $\Xi_{\alpha \beta}^{(l,k)}$ for a plasma with $N$ ion species, which were used to evaluate the dynamic friction and thermal forces through Eqs.~(\ref{eq: dyn-friction}) and (\ref{eq: thermal-force}). Similar expressions for other transport coefficients obtained from existing prescriptions are summarized in Ref.~\cite{kagan2016transport}. Ref.~\cite{kagan2016transport} also provides the numerical routines where these expressions are implemented and which can be used to reproduce the results presented in this paper.

In what follows, $x_{\alpha} = n_{\alpha}/n_i$ and $c_{\alpha} = \rho_{\alpha}/\rho$ denote the number and mass fractions of the ion species $\alpha$, respectively, where $n_{\alpha}$ and $\rho_{\alpha}$ are the number and mass densities of the ion species $\alpha$, respectively, and $n_i = \sum_{\alpha} n_{\alpha}$ and $\rho=\sum_{\alpha} \rho_{\alpha}$ are the total number and mass densities of the ionic mixture, respectively.

\subsection{Matrix representation}
A number of equivalent representations can be found in literature~\cite{hirschfelder1954molecular, devoto1966transport, ferziger1972mathematical, zhdanov2002transport}.  
Here we utilize the formalism by Ferziger and Kaper~\cite{ ferziger1972mathematical}   and use the Kramers rule to write the $\xi$th Chapman-Enskog approximation to the ordinary and thermo-diffusion coefficients in the form of Ref.~\cite{devoto1966transport}
\begin{multline}
[D_{\alpha\beta}]_{\xi} = 
- \frac{4}{25 n_i  |\tensor{M}|} \times  \\
 \label{eq: ordin-diff}
 \begin{vmatrix}
  \tensor{M}^{(0,0)} & \tensor{M}^{(0,1)}  & \ldots   & \tensor{M}^{(0,\xi-1)} &\vec{v}_{k}  \\
  \tensor{M}^{(1,0)} & \tensor{M}^{(1,1)}  & \ldots   & \tensor{M}^{(1,\xi-1)} & \vec{0}  \\
  \vdots                   & \vdots                    & \ddots  & \vdots                        & \vdots \\
  \tensor{M}^{(\xi-1,0)} & \tensor{M}^{(\xi-1,1)} & \ldots   & \tensor{M}^{(\xi-1,\xi-1)} & \vec{0}\\
 \vec{ \delta }_{k\alpha}        & \vec{0}                                & \ldots          &             \vec{0}        & 0
 \end{vmatrix}
\end{multline}
and
\begin{multline}
\label{eq: thermo-diff}
[D_{\alpha}^{(T)}]_{\xi}=
- \frac{2}{5 n_i  |\tensor{M}|}  \times \\
 \begin{vmatrix}
  \tensor{M}^{(0,0)} & \tensor{M}^{(0,1)}  & \ldots   & \tensor{M}^{(0,\xi-1)} & \vec{0} \\
  \tensor{M}^{(1,0)} & \tensor{M}^{(1,1)}  & \ldots   & \tensor{M}^{(1,\xi-1)} & \vec{x}_{k}  \\
  \vdots                   & \vdots                    & \ddots  & \vdots                        & \vdots \\
  \tensor{M}^{(\xi-1,0)} & \tensor{M}^{(\xi-1,1)} & \ldots   & \tensor{M}^{(\xi-1,\xi-1)} & \vec{0}\\
  \vec{ \delta }_{k\alpha}        & \vec{0}                                & \ldots          &               \vec{0}        & 0 
 \end{vmatrix},
\end{multline}where blocks $\tensor{M}^{(i,j)}$ are $N \times N$ matrices, whose elements are provided in the next subsection. In Eqs.~(\ref{eq: ordin-diff}) and (\ref{eq: thermo-diff}),  $|\tensor{M}|$ denotes the  determinant of the $\xi N \times \xi N$ matrix $\tensor{M}$ composed of $\tensor{M}^{(i,j)}$. The  determinants in the numerator are obtained by appending $\tensor{M}$ with a row and a column that are, in turn, composed of  $N$-element vectors indicated by the arrow sign and the last element, scalar 0. The $k$-th element in such a vector is given by the corresponding expressions, in which $\delta_{kl}$ is the Kronecker delta and $v_k$ appearing in the upper right corner in the numerator on the right side of Eq.~(\ref{eq: ordin-diff}) is equal to $0$ for $k=1$ and $\delta_{k\beta} - c_k$ for $2 \leq k \leq N$.

In the employed formalism the ordinary diffusion coefficients are symmetric, $D_{\alpha\beta} = D_{\beta\alpha}$, and also satisfy the constraints 
$\sum_{\alpha} c_{\alpha} D_{\alpha\beta} = 0$ for $\beta = 1..N$, so there are only $N(N-1)/2$ independent coefficients. Thermo-diffusion coefficients satisfy the constraint $\sum_{\alpha} c_{\alpha} D_{\alpha}^{(T)} = 0$, so there are $N-1$ independent coefficients.

\begin{widetext}
\subsection{Matrix elements in terms of the generalized Coulomb logarithms}
\label{app: matrix-elem}

It is convenient to introduce $ \bar{\Xi}_{\alpha \beta}^{(l,k)} \equiv \Xi_{\alpha \beta}^{(l,k)}/\Xi_{\alpha \beta}^{(1,1)}$. Then, elements of matrix $\tensor{M}$ can be written as follows: for the first row of the uppermost leftmost block ($i=j=0,\alpha=1$)
\begin{equation}
\label{eq: M-elem-1}
M_{1\beta}^{(0,0)} = c_{\beta},~ \beta =1..N,
\end{equation}
for the first rows of the remaining uppermost blocks ($i=0, 0<j\leq \xi-1,\alpha=1$)
\begin{equation}
\label{eq: M-elem-1}
M_{1\beta}^{(0,j)} = 0,~ \beta =1..N
\end{equation}
and for all other elements
\begin{equation}
\label{eq: M-elem-2}
M_{\alpha\beta}^{(i,j)} = \frac{8(m_{\alpha}m_{\beta})^{1/2}}{75 T_i}
\Bigl( \delta_{\alpha\beta} \sum_{\chi =1}^{N} x_{\alpha} x_{\chi} A_{\alpha\chi}^{(i,j)} +
x_{\alpha} x_{\beta} B_{\alpha\beta}^{(i,j)} \Bigr),
\end{equation}
where $A_{\alpha\beta}^{(i,j)}$ and $B_{\alpha\beta}^{(i,j)}$ are related to standard bracket integrals~\cite{ ferziger1972mathematical} 
so one can find $A_{\alpha \beta}^{(l,k)} = 3  \nu_{\alpha \beta}/(16  n_{\beta}) \bar{A}_{\alpha \beta}^{(l,k)} $ and $B_{\alpha \beta}^{(l,k)} = 3 \nu_{\alpha \beta}/(16  n_{\beta}) \bar{B}_{\alpha \beta}^{(l,k)} $ with $\nu_{\alpha \beta}$ defined by Eq.~(\ref{eq: nu}) and
\begin{align}
\label{eq: bracket-1}
\bar{A}_{\alpha\beta}^{(0,0)} &= 8 \mu_{\beta}  \\
\bar{A}_{\alpha\beta}^{(0,1)} &= 8 \mu_{\beta}^2  \Bigl(  \frac{5}{2}   -  
\bar{\Xi}_{\alpha \beta}^{(1,2)} \Bigr)\\
\bar{A}_{\alpha\beta}^{(1,1)} &= 8 \mu_{\beta}  \Bigl[\frac{5}{4} (6 \mu_{\alpha}^2 + 5 \mu_{\beta}^2)   -  
5 \mu_{\beta}^2 \bar{\Xi}_{\alpha \beta}^{(1,2)} +
\mu_{\beta}^2 \bar{\Xi}_{\alpha \beta}^{(1,3)} +
2 \mu_{\alpha} \mu_{\beta} \bar{\Xi}_{\alpha \beta}^{(2,2)} \Bigr]\\
\bar{A}_{\alpha\beta}^{(0,2)} &= 4 \mu_{\beta}^3  \Bigl(  \frac{35}{4} -  
7 \bar{\Xi}_{\alpha \beta}^{(1,2)} +
\bar{\Xi}_{\alpha \beta}^{(1,3)} \Bigr)\\
\bar{A}_{\alpha\beta}^{(1,2)} &= 8 \mu_{\beta}^2  \Bigl[\frac{35}{16} (12 \mu_{\alpha}^2 + 5 \mu_{\beta}^2)  -  
\frac{21}{8} (4 \mu_{\alpha}^2 + 5 \mu_{\beta}^2) \bar{\Xi}_{\alpha \beta}^{(1,2)} +
\frac{19}{4} \mu_{\beta}^2 \bar{\Xi}_{\alpha \beta}^{(1,3)}  
- \frac{1}{2} \mu_{\beta}^2 \bar{\Xi}_{\alpha \beta}^{(1,4)} 
+  7 \mu_{\alpha} \mu_{\beta} \bar{\Xi}_{\alpha \beta}^{(2,2)} -
2 \mu_{\alpha} \mu_{\beta} \bar{\Xi}_{\alpha \beta}^{(2,3)} 
\Bigr]\\
\nonumber
\bar{A}_{\alpha\beta}^{(2,2)} &= 8 \mu_{\beta}  \Bigl[\frac{35}{64} (40 \mu_{\alpha}^4 + 168 \mu_{\alpha}^2 \mu_{\beta}^2  +35 \mu_{\beta}^4)  -  
\frac{7}{8} \mu_{\beta}^2 (84 \mu_{\alpha}^2 + 35 \mu_{\beta}^2) \bar{\Xi}_{\alpha \beta}^{(1,2)} 
+  \frac{1}{8} \mu_{\beta}^2 (108 \mu_{\alpha}^2 + 133 \mu_{\beta}^2) \bar{\Xi}_{\alpha \beta}^{(1,3)}  \\
& - \frac{7}{2} \mu_{\beta}^4 \bar{\Xi}_{\alpha \beta}^{(1,4)}+
\frac{1}{4} \mu_{\beta}^4 \bar{\Xi}_{\alpha \beta}^{(1,5)}
+ \frac{7}{2} \mu_{\alpha} \mu_{\beta}  (4 \mu_{\alpha}^2 + 7 \mu_{\beta}^2) \bar{\Xi}_{\alpha \beta}^{(2,2)} - 
14 \mu_{\alpha} \mu_{\beta}^3 \bar{\Xi}_{\alpha \beta}^{(2,3)} +
2 \mu_{\alpha} \mu_{\beta}^3 \bar{\Xi}_{\alpha \beta}^{(2,4)} +
2 \mu_{\alpha}^2 \mu_{\beta}^2 \bar{\Xi}_{\alpha \beta}^{(3,3)}
\Bigr]\\
\bar{B}_{\alpha\beta}^{(0,0)} &= - 8 \mu_{\alpha}^{1/2} \mu_{\beta}^{1/2}  \\
\bar{B}_{\alpha\beta}^{(0,1)} &= - 8 \mu_{\alpha}^{3/2} \mu_{\beta}^{1/2}  \Bigl(  \frac{5}{2}   -  
\bar{\Xi}_{\alpha \beta}^{(1,2)} \Bigr)  \\
\bar{B}_{\alpha\beta}^{(1,1)} &= - 8 \mu_{\alpha}^{3/2} \mu_{\beta}^{3/2}  \Bigl(  \frac{55}{4}   -  
5  \bar{\Xi}_{\alpha \beta}^{(1,2)} +
 \bar{\Xi}_{\alpha \beta}^{(1,3)} -
2 \bar{\Xi}_{\alpha \beta}^{(2,2)} \Bigr)  \\
\bar{B}_{\alpha\beta}^{(0,2)} &=  - 4 \mu_{\alpha}^{5/2} \mu_{\beta}^{1/2}  \Bigl(  \frac{35}{4}   -  
7  \bar{\Xi}_{\alpha \beta}^{(1,2)} +
 \bar{\Xi}_{\alpha \beta}^{(1,3)}  \Bigr)  \\
\bar{B}_{\alpha\beta}^{(1,2)} &= - 8 \mu_{\alpha}^{5/2} \mu_{\beta}^{3/2}  \Bigl(   \frac{595}{16}   -  
\frac{189}{8}  \bar{\Xi}_{\alpha \beta}^{(1,2)} +
\frac{19}{4}  \bar{\Xi}_{\alpha \beta}^{(1,3)} -
\frac{1}{2}  \bar{\Xi}_{\alpha \beta}^{(1,4)} - 
7 \bar{\Xi}_{\alpha \beta}^{(2,2)} +
2 \bar{\Xi}_{\alpha \beta}^{(2,3)} \Bigr)  \\
\nonumber
\bar{B}_{\alpha\beta}^{(2,2)} &= - 8 \mu_{\alpha}^{5/2} \mu_{\beta}^{5/2}  \Bigl(   \frac{8505}{64}   -  
\frac{833}{8}  \bar{\Xi}_{\alpha \beta}^{(1,2)} +
\frac{241}{8}  \bar{\Xi}_{\alpha \beta}^{(1,3)} -
\frac{7}{2}  \bar{\Xi}_{\alpha \beta}^{(1,4)} +
\frac{1}{4}  \bar{\Xi}_{\alpha \beta}^{(1,5)}  
\label{eq: bracket-1-end}
- \frac{77}{2} \bar{\Xi}_{\alpha \beta}^{(2,2)} + 
14 \bar{\Xi}_{\alpha \beta}^{(2,3)} \\ 
& - 2 \bar{\Xi}_{\alpha \beta}^{(2,4)} +
2 \bar{\Xi}_{\alpha \beta}^{(3,3)}  \Bigr).
\end{align}
In Eqs.~(\ref{eq: bracket-1})-(\ref{eq: bracket-1-end}),  $\mu_{\alpha} = m_{\alpha}/(m_{\alpha} + m_{\beta})$ and $\mu_{\beta} = m_{\beta}/(m_{\alpha} + m_{\beta})$ and due to symmetry properties of the bracket integrals
 $\bar{A}_{\alpha\beta}^{(i,j)} = \bar{A}_{\alpha\beta}^{(j,i)}$ and $\bar{B}_{\alpha\beta}^{(i,j)} = \bar{B}_{\beta\alpha}^{(j,i)}$~\cite{ ferziger1972mathematical}. The above expressions for $A_{\alpha\beta}^{(i,j)}$ and $B_{\alpha\beta}^{(i,j)}$ with $i,j \leq 2$ are thus sufficient for evaluating  the first to third order Chapman-Enskog approximations to the transport coefficients.

\end{widetext}

\section{Equation for the concentration evolution}
\label{app: concentration-evolution}

In this paper we consider the ionic mixture only, electrons are assumed to provide the neutralizing background. Then, the complete expression for the diffusive flux of the light ion species has the form
\begin{equation}
\label{eq: canonical-flux}
\vec{i} =
- \rho D \Bigl( \nabla c +k_p \nabla \log{p_i} + \frac{e k_E}{T_i}\nabla \Phi \Bigr) 
+ D_T \nabla \log{T_i} ,
\end{equation}
where $\Phi$ is the electrostatic potential accounting for the ambipolar field and $p$ is the mixture pressure. The Landau and Lifshitz's ``classical diffusion coefficient'' $D$~\cite{landau1987fluid} on the right side of Eq.~\ref{eq: canonical-flux} is related to the ordinary diffusion coefficient  $D_{12}$ of Ferziger and Kaper~\cite{ferziger1972mathematical}, employed in the present work,  through 
\begin{equation}
\label{eq: class-diff}
D = - \frac{\rho^2}{m_1 m_2 n_i^2} D_{12}
\end{equation}
and is also equal to the ``binary diffusion coefficient'' $\mathcal{D}_{12}$ of Ferziger and Kaper. 

On the right side of Eq.~\ref{eq: canonical-flux}, the so-called baro- and electro-diffusion ratios, $k_p$ and $k_E$,  can be expressed through derivatives of the electro-chemical potential of the mixture~\cite{landau1987fluid, kagan2014thermodynamic}. For non-ideal plasmas, their calculation therefore requires a separate, thermodynamic analysis which is not presented in this paper.

The kinetic analysis conducted in this paper is sufficient for the main conclusions such as that thermo-diffusion diminishes for substantial ion couplings. For practical modeling of multi-component plasmas one would also need the thermodynamic terms. Then, the diffusive flux~(\ref{eq: canonical-flux}) can be evaluated from the hydrodynamic variables and species concentration evolved with
\begin{equation}
\label{eq:dcdt}
\rho \frac{\partial c}{\partial t} + \rho \vec{u}  \cdot\nabla c + \nabla\cdot \vec{i} = 0,
\end{equation}
 where $\vec{u} $ is the center-of-mass velocity. 


\end{document}